\documentclass[11pt]{article}

\usepackage[preprint]{acl}

\usepackage{times}
\usepackage{latexsym}

\usepackage[T1]{fontenc}
\usepackage[utf8]{inputenc}

\usepackage{microtype}

\usepackage{inconsolata}

\usepackage{graphicx}
\usepackage[table,xcdraw]{xcolor}
\usepackage{multirow}
\usepackage{float}
\usepackage{booktabs}

\usepackage{pifont}
\newcommand{\yes}{\textcolor{black!60!black}{\ding{52}}} % 52是加粗对勾，51是普通对勾
\newcommand{\no}{\textcolor{black!60!black}{\ding{56}}} % 56是加粗叉号，55是普通叉号

\usepackage[most]{tcolorbox}
\usepackage{xcolor}
\usepackage{enumitem} % 用于调整列表间距

\newtcolorbox{promptbox}[2][]{
  colback=white,            % 背景纯白
  boxrule=0.8pt,            % 边框宽度
  arc=2mm,                  % 圆角
  fonttitle=\bfseries\sffamily, % 标题字体
  fontupper=\small\ttfamily,    % 内容字体：这里统一设定了 \small
  left=3mm, right=3mm, top=3mm, bottom=3mm, % 内边距
  title={#2},               % 标题
  #1                        % 允许传入颜色等额外参数
}
\definecolor{exframe}{RGB}{80, 80, 80}   % 深灰边框
\definecolor{exback}{RGB}{250, 250, 250} % 极淡灰背景

\newtcolorbox{fulltaskbox}[1][]{
  enhanced,
  breakable,                % 允许跨页，这是核心
  colback=exback,           % 背景色
  colframe=exframe,         % 边框色
  coltitle=white,
  fonttitle=\bfseries\sffamily,
  title={Dataset Samples},  % 默认标题
  boxrule=0.8pt,
  arc=2mm,
  left=3mm, right=3mm, top=3mm, bottom=3mm,
  attach boxed title to top left={xshift=4mm, yshift*=-\tcboxedtitleheight/2},
  boxed title style={
    colback=exframe,
    frame hidden,
    sharp corners,
    top=1mm, bottom=1mm, left=2mm, right=2mm
  },
  #1 % 允许修改参数
}

\newtcolorbox{failurecase}[2][]{
  colback=gray!5!white,
  colframe=black!75!white,
  fonttitle=\bfseries,
  title={#2},
  #1
}

\title{GenomeQA: Benchmarking General Large Language Models for Genome Sequence Understanding}

\author{
 \textbf{Weicai Long\textsuperscript{1,$\dagger$}},
 \textbf{Yusen Hou\textsuperscript{1,$\dagger$}},
 \textbf{Junning Feng\textsuperscript{1}},
 \textbf{Houcheng Su\textsuperscript{1}},
\\
 \textbf{Shuo Yang\textsuperscript{2}},
 \textbf{Donglin Xie\textsuperscript{3}},
 \textbf{Yanlin Zhang\textsuperscript{1,*}},
\\
 \textsuperscript{1}Hong Kong University of Science and Technology (Guangzhou), \\
 \textsuperscript{2}The University of Hong Kong, \\
 \textsuperscript{3}Peking University
\\
 \small{
   \textbf{Correspondence*:} \href{yanlinzhang@hkust-gz.edu.cn}{yanlinzhang@hkust-gz.edu.cn}
 }
}

\begin{document}
\maketitle

\begingroup
\renewcommand\thefootnote{$\dagger$}
\footnotetext{Co-first authors}
\endgroup

\begin{abstract}
Large Language Models (LLMs) are increasingly adopted as conversational assistants in genomics, where they are mainly used to reason over biological knowledge, annotations, and analysis outputs through natural language interfaces. However, existing benchmarks either focus on specialized DNA models trained for sequence prediction or evaluate biological knowledge using text-only questions, leaving the behavior of general-purpose LLMs when directly exposed to raw genome sequences underexplored. We introduce GenomeQA, a benchmark designed to provide a controlled evaluation setting for general-purpose LLMs on sequence-based genome inference tasks. GenomeQA comprises 5,200 samples drawn from multiple biological databases, with sequence lengths ranging from 6 to 1,000 base pairs (bp), spanning six task families: Enhancer and Promoter Identification, Splice Site Identification, Taxonomic Classification, Histone Mark Prediction, Transcription Factor Binding Site Prediction, and TF Motif Prediction. Across six frontier LLMs, we find that models consistently outperform random baselines and can exploit local sequence signals such as GC content and short motifs, while performance degrades on tasks that require more indirect or multi-step inference over sequence patterns. GenomeQA establishes a diagnostic benchmark for studying and improving the use of general-purpose LLMs on raw genomic sequences\footnote{Data and code are available at https://anonymous.4open.science/r/GenomeQA-E350}.
\end{abstract}

\section{Introduction}

\begin{figure*}[h]
  \includegraphics[width=\linewidth]{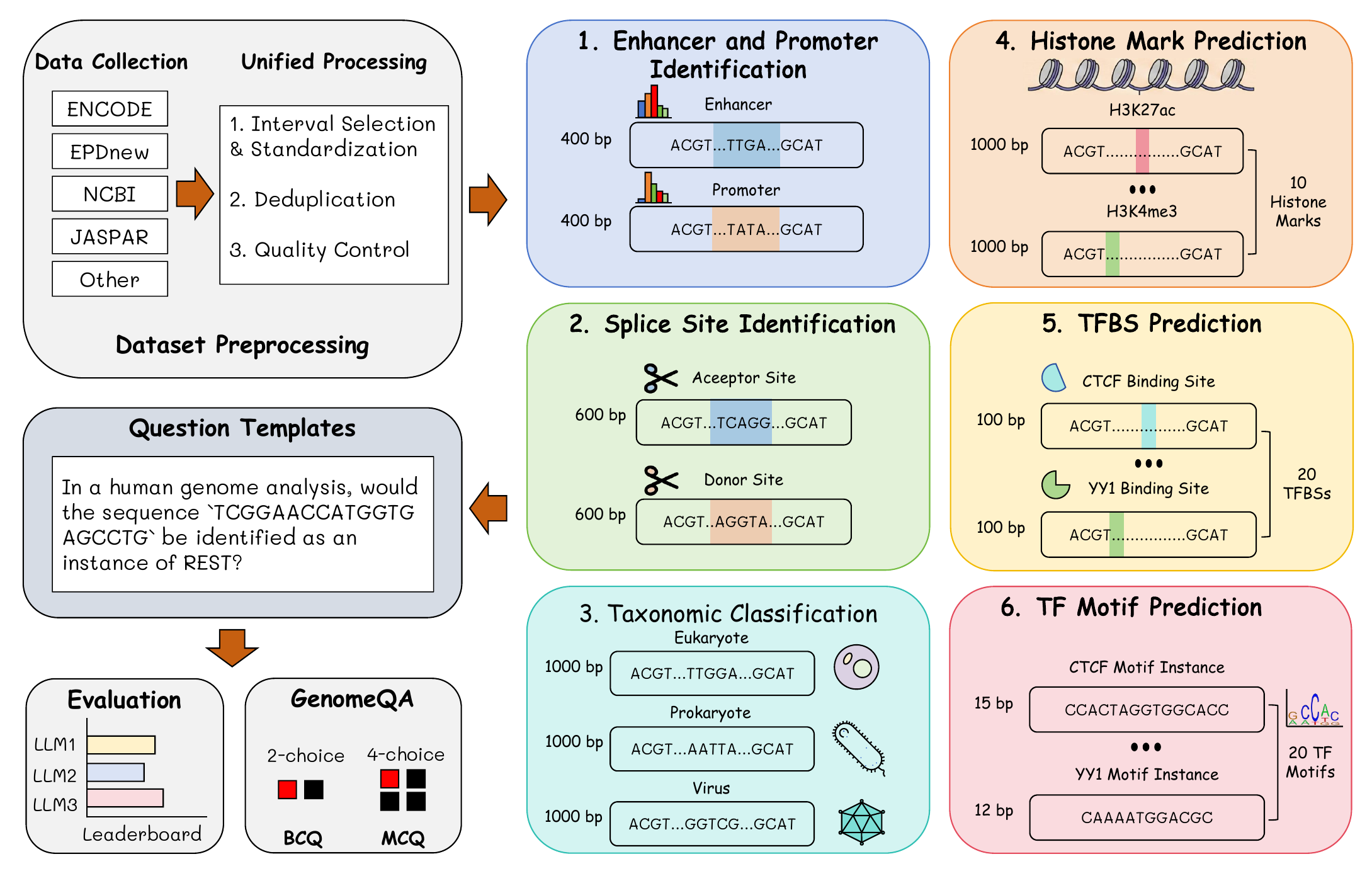}
  \caption{Overview of GenomeQA. The pipeline consists of Data collection, processing, construction and LLM evaluation.}
  \label{fig:overview}
\end{figure*}

% place after 'In genomics xxx': More recently, multimodal systems have begun to connect LLMs with genomic pipelines or sequence encoders, enabling interactive querying over sequence-derived signals rather than requiring the LLM to directly model raw DNA strings.

Genome analysis has long relied on specialized sequence models and task-specific pipelines. Recent years have seen rapid progress in DNA foundation models that are trained directly on nucleotide sequences, as well as emerging DNA-Text systems that couple a dedicated DNA encoder with a large language model \citep{zhou2023dnabert2, dalla2025nt, brixi2025evo2, schiff2024caduceus, de2025chatnt, fallahpour2025bioreason}. These approaches achieve strong supervised performance, but they typically require task-specific heads, probing, or additional adaptation to support downstream applications. 
In parallel, large language models have become widely used as conversational assistants across scientific domains, including chemistry \citep{llm_chemistry, llm_chemcrow}, physics \citep{llm_physics1, llm_physics2}, and medicine \citep{llm_medicine1, llm_medicine2}. In genomics, their most common role is to reason over derived information such as annotations for genes and variants, functional summaries, experimental metadata, literature, and results from existing bioinformatics tools through natural language interfaces. 
However, direct interaction with raw genomic sequences poses challenges that differ fundamentally from standard natural language processing. DNA lacks human-interpretable semantic units such as words or grammar, exhibits long-range dependencies, and encodes biological signals in highly degenerate and context-dependent patterns \citep{Cheng2025dnalongbench, gpn_msa}. As a result, it remains unclear how general-purpose LLMs behave when they are directly exposed to nucleotide sequences, and whether their responses reflect non-trivial sequence-level cues or are driven primarily by superficial heuristics.

Despite the increasing presence of LLMs in genomics studies, there is currently no standardized benchmark that evaluates how general-purpose language models perform when directly exposed to raw genomic sequences. Existing genome benchmarks \citep{marin2024bend, genomic_benchmarks, patel2024darteval} primarily target trainable DNA-specific models and assess representation quality under fine-tuning or probing. In contrast, exam-style benchmarks \citep{chen2025domain_benchmark, queen2025cgbench, genome_bench} evaluate biological knowledge using text-only questions without requiring sequence-level analysis. Consequently, a practically important evaluation setting remains underexplored: a general-purpose LLM receiving natural language questions together with raw nucleotide sequences and producing answers under a fixed instruction-following protocol.

We take a step toward filling this gap by introducing GenomeQA, a question-answering benchmark designed to provide a controlled evaluation of general-purpose LLMs on sequence-based genome inference tasks. As shown in Figure~\ref{fig:overview}, GenomeQA comprises 5,200 samples across six representative task families: Enhancer and Promoter Identification, Splice Site Identification, Taxonomic Classification, Histone Mark Prediction, Transcription Factor Binding Site (TFBS) Prediction, and Transcription Factor (TF) Motif Prediction. Each instance is formatted as either a Binary Choice Question (BCQ) for validity judgment or a four-option Multiple Choice Question (MCQ). To ensure consistent evaluation across models and tasks, we use a single system prompt derived from a small pilot study; the prompt is fixed throughout all experiments and provides domain-relevant guidance for analyzing sequence signals such as motifs and base composition.

Our contributions are summarized as follows:
\begin{itemize}
    \item We introduce GenomeQA, a benchmark that provides a controlled evaluation setting for assessing how general-purpose LLMs perform on sequence-based genome inference tasks. The benchmark comprises 5,200 curated samples spanning six biologically grounded task families.
    \item We conduct a comprehensive evaluation of six frontier LLMs, establishing baseline performance on raw DNA sequences and showing that current models can exploit certain local sequence signals (e.g., GC content and short motifs) but struggle with tasks requiring more complex or indirect inference.
    \item We present a fine-grained analysis of failure modes such as Sequence Motif Over-reliance and Character Fidelity Loss, highlighting systematic error patterns and directions for future research.
\end{itemize}

\section{Related Work}
\subsection{Genome-Specific Models}

Recent DNA foundation models treat genome sequences as language. They pretrain on billions of nucleotides and add task-specific heads for downstream applications such as regulatory elements prediction and splice site identification. Typical models include DNABERT-2\citep{zhou2023dnabert2}, Nucleotide Transformer\citep{dalla2025nt}, HyenaDNA\citep{nguyen2023hyenadna}, Genos\citep{genos}, GENA-LM\citep{gena_lm} and Evo\citep{evo1}. 
Recent multimodal approaches couple a pretrained DNA foundation model with a general-purpose large language model (LLM), such as ChatNT\citep{de2025chatnt}, BioReason\citep{fallahpour2025bioreason} and Omni-DNA\citep{li2025omnidna}. Empirically, these models achieve strong performance and can match or even surpass state-of-the-art results on standard genome understanding benchmarks. Meanwhile, some researchers argue that an alternative route is to repurpose general LLMs into DNA-LLMs directly\citep{cheng2025l2g}, e.g., by adapting tokenization and training objectives so that the LLM can model DNA sequences without an explicit separate DNA encoder. Overall, despite their different interfaces, these methods share a key property: they rely on models that are specifically trained or substantially adapted on DNA sequences.

\subsection{Genome Benchmarks}

Existing benchmarks for genome modeling primarily target models that are specifically designed or adapted for genome data. Sequence-based benchmarks, such as BEND \citep{marin2024bend}, DNALongBench \citep{Cheng2025dnalongbench}, DART-Eval \citep{patel2024darteval}, and the Genomics Long-range Benchmark \citep{genomic_lr_bench}, are mainly constructed to evaluate genome foundation models. These benchmarks typically consist of raw DNA sequences paired with predefined prediction tasks, and are intended to measure a model’s ability to learn transferable sequence representations for downstream biological applications. Another line of benchmarks focuses on biomedical or clinical knowledge assessment through natural language questions, such as CMExam \citep{cmexam}, Bio-Benchmark \citep{bio_benchmark}, and EHRXQA \citep{ehrxqa}. While these benchmarks evaluate language understanding and domain knowledge, they do not require models to directly process or reason over raw genomic sequences. More recently, several benchmarks and evaluation protocols have emerged in the context of multimodal genome-language systems, including Lab-Bench \citep{laurent2024lab_bench} and the task suites used in ChatNT and BioReason. These benchmarks partially involve sequence-based question answering, but they rely on a dedicated genome encoder to transform DNA sequences into latent representations before passing them to a large language model. As a result, the evaluated capability is that of an integrated multimodal system, rather than the intrinsic ability of a general-purpose LLM to interpret DNA sequences. In contrast, GenomeQA is designed to assess general-purpose LLMs in a setting where raw DNA sequences are provided directly as input. Each instance contains real genomic sequences with task-relevant signals, and all tasks are reformulated into a unified natural language question-answering format. GenomeQA does not assume a genome-specific encoder or additional training on DNA data, and is intended to support controlled measurement of LLMs performance on sequence-based genome inference tasks.

\section{GenomeQA Construction}
\subsection{Design Principles}
\label{sec:design_principles}

% Please add the following required packages to your document preamble:
% \usepackage{booktabs}
% \usepackage{multirow}
% \usepackage{graphicx}
\begin{table*}[]
\centering
\resizebox{\textwidth}{!}{%
\begin{tabular}{@{}cccccccc@{}}
\toprule
\multirow{2}{*}{\textbf{Task}} & \multirow{2}{*}{\textbf{Source}} & \multirow{2}{*}{\textbf{Seq. Len}} & \multirow{2}{*}{\textbf{Label}} & \multicolumn{2}{c}{\textbf{BCQ}} & \multicolumn{2}{c}{\textbf{MCQ}} \\ \cmidrule(l){5-8} 
 &  &  &  & \textbf{Counts} & \textbf{Avg. Len} & \textbf{Counts} & \textbf{Avg. Len} \\ \midrule
Enhancer and Promoter Identification & EPDnew,SCREEN & 400 & 2 & 500 & 176 & 500 & 70 \\
Splice Site Identification & NT downstream tasks & 600 & 3 & 500 & 254 & 500 & 94 \\
Taxonomic Classification & NCBI RefSeq & 1000 & 3 & 500 & 402 & 500 & 146 \\
Histone Mark Prediction & NT downstream tasks & 1000 & 10 & 500 & 419 & 500 & 176 \\
TFBS Prediction & ENCODE, JASPAR & 100 & 20 & 500 & 61 & 500 & 31 \\
TF Motif Prediction & ENCODE, JASPAR & 6-20 & 20 & 100 & 23 & 100 & 16 \\ \bottomrule
\end{tabular}%
}
\caption{The statistics of GenomeQA suite, where Source indicates the data source, Seq. Len refs to the length of DNA sequence, and Label denotes the number of label sets. Avg. Len represents the average lengths of the questions. We report the number of tokens after tokenization using the Llama-4 tokenizer.}
\label{tab:bench_overview}
\end{table*}

%An overview of Dataset construction is shown in Figure \ref{fig:overview}. The benchmark follows a systematic pipeline designed to transform genome annotations into a rigorous evaluation framework. (1) We first curate six fundamental task families by sourcing high-quality data from open-source databases or repositories such as ENCODE\citep{encode}, EPDnew\citep{dreos2013epdnew}, NCBI\citep{ncbi_online}, JASPAR\citep{jaspar} and Nucleotide Transformer (NT) downstream tasks\citep{dalla2025nt}. A core design principle of this selection is the maintenance of biological hierarchy: we group tasks to compare patterns within the same functional level, such as distinguishing enhancers from promoters, rather than conflating features across disparate biological scales. (2) These tasks are processed through a unified three-stage workflow to ensure data integrity. First, we perform interval selection and standardization to calibrate sequence lengths, ensuring that every window sufficiently encompasses the necessary motifs for identification. Notably, we use Bedtools\citep{bedtools} to extract the sequences. Second, we implement deduplication and overlap resolution to eliminate conflicting annotations and mitigate task ambiguity, thereby enhancing the overall reliability of the dataset. Finally, a quality control filter is applied to exclude sequences containing ambiguous bases that might introduce technical noise. 

An overview of the dataset construction pipeline is shown in Figure~\ref{fig:overview}. GenomeQA is built through a systematic process that transforms curated genome annotations into a unified evaluation framework. (1) We first identify six fundamental task families by collecting high-quality annotations from established databases and repositories, including ENCODE~\citep{encode}, EPDnew~\citep{dreos2013epdnew}, NCBI~\citep{ncbi_online}, JASPAR~\citep{jaspar}, as well as downstream tasks used in the Nucleotide Transformer (NT) benchmark~\citep{dalla2025nt}. 
A central design principle is the preservation of biological hierarchy: tasks are grouped to contrast genomic elements at the same functional level (e.g., enhancers versus promoters), rather than mixing signals across disparate biological scales. (2) All selected tasks are processed through a unified three-stage workflow to ensure consistency and data quality. First, we perform interval selection and length standardization to calibrate sequence windows, ensuring that each sequence sufficiently covers the motifs or regulatory signals required for the task. Sequence extraction is conducted using Bedtools~\citep{bedtools}. Second, we remove overlapping samples in six tasks to reduce ambiguity and improve label reliability. Third, a quality control filter is applied to exclude sequences containing ambiguous nucleotide bases.
(3) This process concludes with a question formulation stage, where each validated DNA sequence is instantiated into standardized natural language templates. Example question templates are provided in Appendix~\ref{sec:question_template}. All tasks are presented in either Binary Choice Question (BCQ) or Multiple Choice Question (MCQ) formats. By constraining the answer space and standardizing the question structure, GenomeQA provides a controlled evaluation setting for assessing models’ ability to reason over raw DNA sequences.

\subsection{Task Families}
\label{sec:task_families}

\textbf{Enhancer and Promoter Identification.} 
This task focuses on distinguishing cis-regulatory elements in the human genome, specifically promoters and enhancers. Promoter candidates are sourced from EPDnew\citep{dreos2013epdnew}, covering regions from 299 base pairs (bp) upstream to 100 bp downstream of the transcription start site. Enhancer candidates are selected from the ENCODE SCREEN database\citep{encode}, with each region centered and resized to match the length of promoter sequences. To ensure that each 400-bp segment corresponds to a single regulatory element, we remove overlapping regions both within and across datasets. The label sets and label statistics are provided in Section~\ref{sec:suppl_task1}.

\textbf{Splice Site Identification.} 
This task focuses on human splice acceptor and donor sites. We aggregate positive 600 bp windows from the NT downstream tasks into a unified pool. After that, we relabel windows overlapping by more than 350 bp as containing both elements, while discarding those with shorter overlaps. For questions about the presence of splice sites, we introduce composition-matched negatives by generating dinucleotide-preserving shuffled controls\citep{dinuc_shuffle} at construction period, forcing models to rely on higher-order motifs rather than simple base composition cues. The label sets and label statistics are provided in \ref{sec:suppl_task2}.

\textbf{Taxonomic Classification.} This task evaluates whether LLMs can recover broad taxonomic groups from sequence alone. We sample 1 kbp fragments from NCBI RefSeq\citep{ncbi_refseq} assemblies representing eukaryote, prokaryote, and virus. Each fragment is assigned a high-level label with approximately balanced sampling across three groups. The complete species list and label statistics are provided in \ref{sec:suppl_task3}.

\textbf{Histone Mark Prediction.} This task focuses on the identification of specific histone modifications from genome sequences in human K562 cells. We utilize 10 distinct histone marks from NT downstream tasks, retaining only positive 1 kbp windows. To ensure unique classification, we remove overlapping windows so that each region is associated with exactly one label. Beyond identifying individual marks , we further assess the model’s understanding of functional chromatin states by categorizing some marks as either open(e.g., H3K4me3, H3K27ac, H3K9ac) or repressive(e.g., H3K9me3, H3K27me3). This multi-dimensional annotation enables us to construct questions ranging from specific mark classification to broader chromatin accessibility that test whether the model perceives the underlying functional similarities between different histone modifications. The full histone mark list and label statistics are provided in \ref{sec:suppl_task4}.

\textbf{TFBS Prediction.} We design this task to evaluate whether LLMs can accurately identify the specific transcription factor binding sites (TFBS) present in a 100 bp genome window. We select 20 transcription factors (TF) with distinct motifs and source their ChIP-seq peaks from ENCODE. For each TF, we select peak intervals by placing a 100 bp window around the summit and then remove any sample pairs that overlap by more than 70 bp. To obtain precise labels, we scan the remaining sequences with FIMO\citep{meme_fimo} using JASPAR position weight matrices (PWM), recording all factors with motif instances in each window. This multi-label dataset enables questions regarding the presence of specific TFs. Additionally, we use CTCF as a canonical architectural protein that organizes chromatin loops and topologically associating domain (TAD) boundaries, constructing questions that test whether models understand its link to 3D genome architecture without naming the factor explicitly. The complete TF list and label statistics are provided in \ref{sec:suppl_task5}.

\textbf{TF Motif Identification.} The final task focuses on short motif instances underlying the TFBS prediction task. Using the FIMO results described above, we collect all motif instances for the same 20 transcription factors and deduplicate them, yielding segments from 6 to 20 bp, each associated with a single TF label. Although the questions are linguistically simple, this task probes whether LLMs encode any recognizable representation of canonical transcription factor motifs. The label sets and label statistics are provided in \ref{sec:suppl_task6}.

\subsection{Dataset statistics} 

\begin{figure}[h]
  \includegraphics[width=\linewidth]{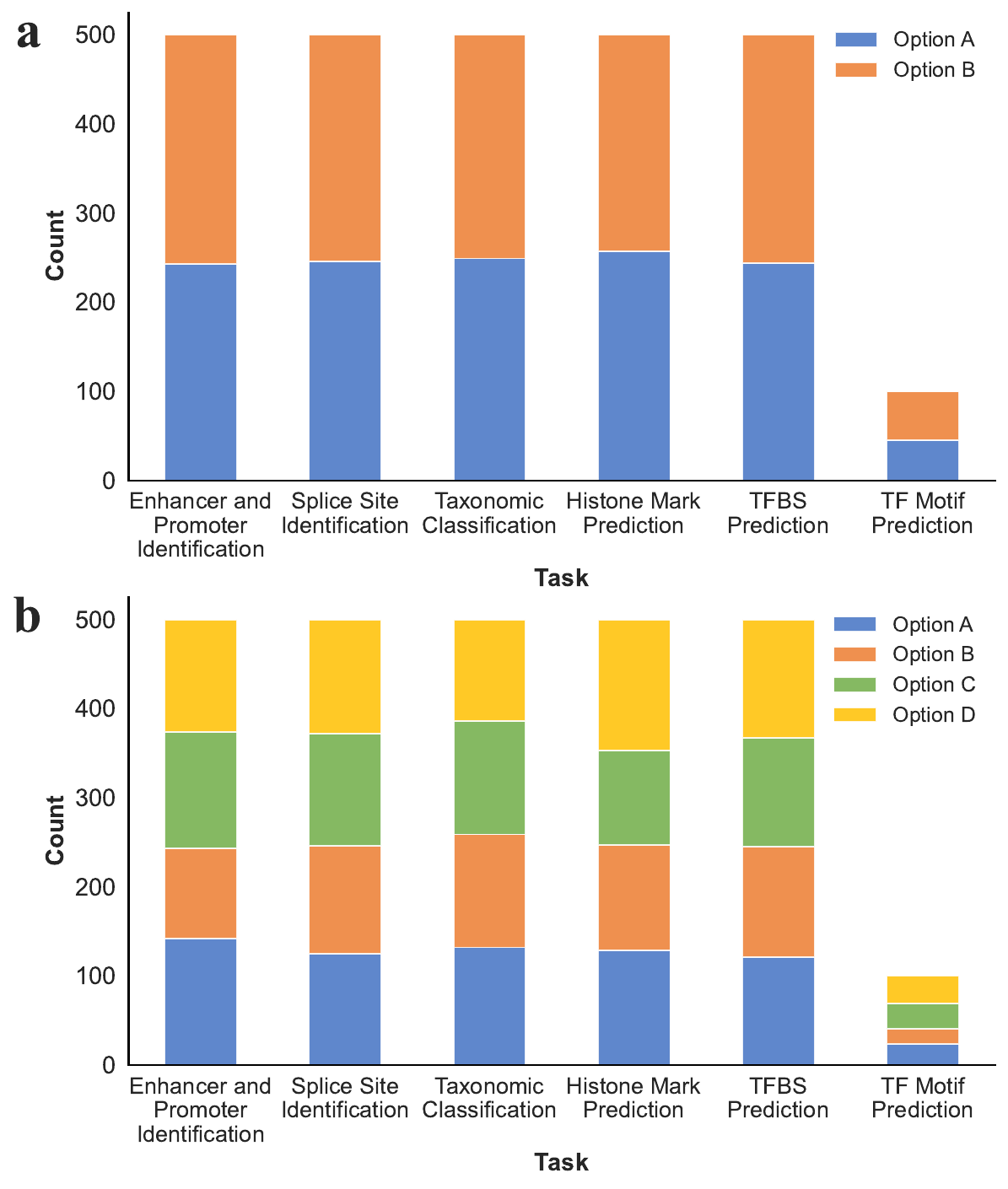}
  \caption{Distribution of Options in BCQ (a) and MCQ (b) of GenomeQA.}
  \label{fig:option_dist}
\end{figure}

Table \ref{tab:bench_overview} provides a comprehensive overview of GenomeQA, detailing the biological focus, data sources, sequence lengths, label set sizes, volumes and average length of question instances for each task. Most tasks contribute 500 BCQs and 500 MCQs, while the simpler motif task includes 100 of each. Sequence lengths range from short motifs (6–20 bp) to medium regulatory windows (100–400 bp) and large genome contexts (1 kbp), requiring models to process both local patterns and broader organizational structures. Figure~\ref{fig:option_dist} displays the distribution of correct answers across option positions for all tasks. These distributions are nearly uniform, confirming that answer keys are balanced and free from position bias. Together, these metrics indicate that GenomeQA is structurally balanced, ensuring that evaluation results reflect genuine sequence understanding rather than dataset artifacts.

\section{Experiments and Analysis}

\subsection{Experimental setup}
% We excluded DeepSeek V3.2 and Kimi K2 due to prohibitive inference latency during reasoning.

We describe the evaluated models, prompt configuration, and evaluation metrics below.

\textbf{Baseline Models.} We evaluate six state-of-the-art general large language models to assess their capabilities on genome data. The model set includes proprietary frontier models: Claude-Sonnet-4.5\citep{claude_sonnet_4_5}\footnote{Developed by Anthropic.}, GPT-5.1\citep{gpt_5_1}\footnote{Developed by OpenAI.}, Gemini-3-Pro\citep{gemini_3_pro}\footnote{Developed by Google. We use Gemini-3-Pro-Preview.}, Grok-4.1\citep{grok_4_1}\footnote{Developed by xAI. We use Grok-4.1-Fast.}, Llama-4\citep{llama4}\footnote{Developed by Meta. We use Llama-4-Maverick-17B-128E-Instruct.}, Qwen3-Max\citep{qwen3_max}\footnote{Developed by Alibaba. We use Qwen3-Max-Preview.}. We enable thinking mode whenever supported for all models to maximize their potential for complex biological deduction.

\textbf{Prompt Settings.} We use a single fixed system prompt across all tasks and models, with details provided in Appendix~\ref{sec:prompt_ablation}. No task-specific few-shot examples are included. The prompt provides domain-relevant guidance and a standardized output format for analyzing sequence signals (e.g., motifs and base composition), ensuring consistency across models while minimizing variation due to prompt design.

\textbf{Metrics.} We select classification accuracy as the evaluation metric. Since the option letters are randomly permuted and correspond to raw DNA sequences rather than stable semantic labels, we calculate accuracy by comparing the model-selected option letter against the ground truth.

% We evaluate 6 state-of-the-art general-purpose language models: Claude-Sonnet-4.5, GPT-5.1, Gemini-3-Pro, Grok-4.1, Llama-4, and Qwen3-Max. For the included models, we enable reasoning capabilities whenever supported to maximize performance. All models share a single domain-aware system prompt. We design this prompt using a small set of questions and the detailed information are shown in Appendix \ref{sec:prompt_ablation}. After selecting the best version, we freeze it and use it for all tasks and models without providing task-specific examples. We report accuracy as the primary metric. Furthermore, because option letters are randomly permuted and often correspond to raw DNA sequences rather than stable semantic labels, computing F1 over option positions would primarily capture positional bias rather than biological discrimination. Consequently, we rely on accuracy for all main results.

\subsection{Main Results}

% Please add the following required packages to your document preamble:
% \usepackage{graphicx}
% \usepackage[table,xcdraw]{xcolor}
% Beamer presentation requires \usepackage{colortbl} instead of \usepackage[table,xcdraw]{xcolor}
\begin{table*}[]
\centering
\resizebox{\textwidth}{!}{%
\begin{tabular}{ccccccccc}
\toprule
\textbf{Model} & \textbf{Thinking} & \textbf{\begin{tabular}[c]{@{}c@{}}Enhancer and Promoter\\ Identification\end{tabular}} & \textbf{\begin{tabular}[c]{@{}c@{}}Splice Site\\ Identification\end{tabular}} & \textbf{\begin{tabular}[c]{@{}c@{}}Taxonomic\\ Classification\end{tabular}} & \textbf{\begin{tabular}[c]{@{}c@{}}Histone Mark\\ Prediction\end{tabular}} & \textbf{\begin{tabular}[c]{@{}c@{}}TFBS\\ Prediction\end{tabular}} & \textbf{\begin{tabular}[c]{@{}c@{}}TF Motif\\ Prediction\end{tabular}} & \textbf{Avg.} \\ \midrule
\multicolumn{9}{c}{\cellcolor[HTML]{FADADE}\textbf{BCQ}} \\ \midrule
Claude-Sonnet-4.5 & \yes & 69.00 & 53.20 & 61.60 & 53.80 & 56.60 & 64.00 & 59.70 \\
GPT-5.1 & \yes & \textbf{70.60} & \textbf{54.40} & 64.60 & 53.20 & 55.80 & 66.00 & 60.77 \\
Gemini-3-Pro & \yes & 67.80 & 50.00 & \textbf{78.80} & \textbf{57.20} & \textbf{59.80} & \textbf{84.00} & \textbf{66.27} \\
Grok-4.1 & \yes & 67.60 & 47.60 & 63.80 & 53.20 & 54.60 & 69.00 & 59.30 \\
Llama-4 & \no & 55.40 & 50.00 & 57.00 & 52.60 & 51.60 & 73.00 & 56.60 \\
Qwen3-Max & \yes & 60.60 & 49.80 & 60.60 & 51.20 & 54.00 & 63.00 & 56.53 \\ \midrule
Random & \no & \multicolumn{7}{c}{50.00} \\ \midrule
\multicolumn{9}{c}{\cellcolor[HTML]{FADADE}\textbf{MCQ}} \\ \midrule
Claude-Sonnet-4.5 & \yes & 59.00 & 28.40 & 64.60 & 35.80 & 48.00 & 81.00 & 52.80 \\
GPT-5.1 & \yes & 62.60 & 26.80 & 61.00 & 36.20 & 50.20 & 77.00 & 52.30 \\
Gemini-3-Pro & \yes & \textbf{66.00} & \textbf{33.40} & \textbf{77.00} & \textbf{37.80} & \textbf{59.00} & \textbf{92.00} & \textbf{60.87} \\
Grok-4.1 & \yes & 56.20 & 25.80 & 61.20 & 33.40 & 48.40 & 79.00 & 50.67 \\
Llama-4 & \no & 36.00 & 26.00 & 56.20 & 31.60 & 34.40 & 65.00 & 41.53 \\
Qwen3-Max & \yes & 43.00 & 25.80 & 52.80 & 33.20 & 45.20 & 72.00 & 45.33 \\ \midrule
Random & \no & \multicolumn{7}{c}{25.00} \\ \bottomrule
\end{tabular}%
}
\caption{Overall results of LLMs on GenomeQA. Thinking denotes whether the model utilizes Chain-of-Thought reasoning. The table reports the classification accuracy (\%) for each subtask and Avg. denotes the average accuracy. The best performance in each task is bolded.}
\label{tab:overall_res}
\end{table*}

% \textbf{(2)Performance correlates strongly with the salience of sequence features.} Model performance varies significantly depending on the task. LLMs achieve respectable accuracy on Enhancer and Promoter Identification, Taxonomic Classification, and TF Motif Prediction. These tasks involve distinct sequence patterns, such as conserved consensus sequences, which are easier to recognize. Conversely, models perform poorly on Splice Site Identification, Histone Mark Prediction, and TFBS Prediction. For these tasks, we explicitly introduce indirect reasoning to increase difficulty. These questions require the model to identify the target intent first and then scan for relevant features. The consistently low accuracy on these tasks proves that current LLMs struggle significantly with this indirect reasoning.

\begin{figure}[h]
  \centering
  \includegraphics[width=0.9\linewidth]{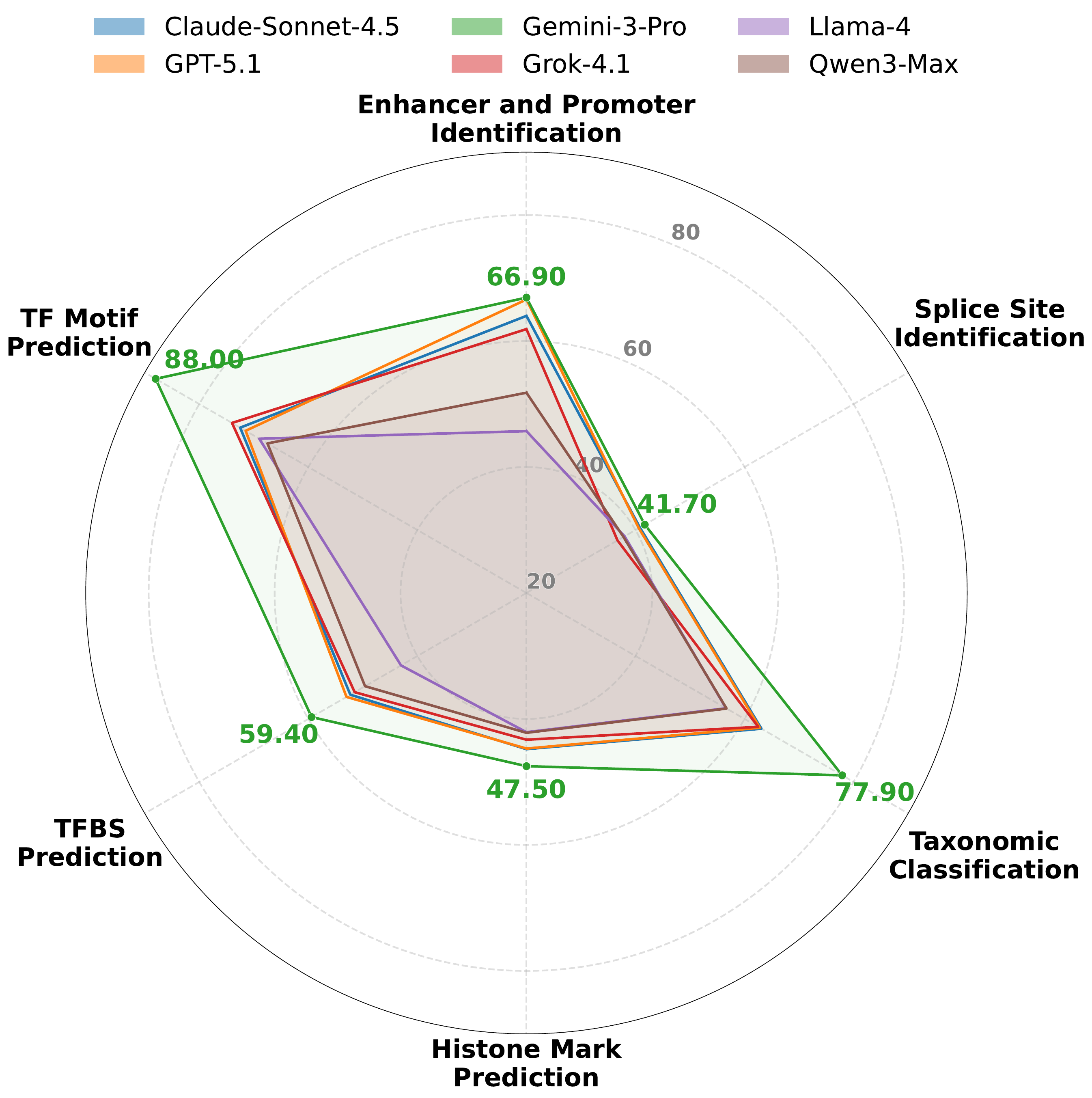}
  \caption{Performance comparison of LLMs across six tasks. The scores represent the macro-average accuracy (\%) derived from both BCQ and MCQ. The numerical annotations indicate the highest accuracy achieved for each task (dominated by Gemini-3-Pro).}
  \label{fig:result_radar}
\end{figure}

Table~\ref{tab:overall_res} and Figure~\ref{fig:result_radar} summarize the performance of the evaluated models on GenomeQA under BCQ and MCQ settings. We highlight three observations. \textbf{(1) Frontier LLMs outperform random baselines but show substantial performance variation across tasks.} 
Among the evaluated models, Gemini-3-Pro achieves the highest average accuracy (66.27\% on BCQ and 60.87\% on MCQ), with Claude-Sonnet-4.5, GPT-5.1, and Grok-4.1 forming a closely clustered second tier. In contrast, Llama-4 and Qwen3-Max exhibit lower overall accuracy (e.g., Qwen3-Max achieves 56.87\% on BCQ and 41.38\% on MCQ). These results indicate that, even for the strongest models, performance on GenomeQA remains weak and inconsistent across tasks. \textbf{(2) Model performance correlates with task complexity and the depth of reasoning required for each task}.
As shown in Figure \ref{fig:result_radar}, Large language models achieve respectable accuracy on Enhancer and Promoter Identification, Taxonomic Classification, and TF Motif Prediction as these questions are formatted as direct pattern recognition. In contrast, performance drops significantly on Splice Site Identification, Histone Mark Prediction, and TFBS Prediction. These difficult tasks involve more complex genome signals such as the long-range patterns of histone marks. Furthermore, the framework incorporates indirect reasoning to increase the complexity of these specific tasks. The consistently low accuracy across these categories proves that current LLMs struggle to execute indirect reasoning when they encounter intricate genome data.  \textbf{(3) Multiple choice formats enhance relative discrimination.} The random baselines are 50.00\% for BCQ and 25.00\% for MCQ, with empirical validation provided in Appendix \ref{suppl_sec:emp_random}. Although absolute accuracy is naturally lower in the multiple choice setting, the relative improvement over the baseline is substantially higher. This trend occurs as the format shifts the task from absolute verification to comparative ranking. Unlike isolated binary decisions, the provided options serve as contextual anchors that narrow the search space, enabling models to evaluate relative likelihoods among candidates. Consequently, the performance gain over chance is approximately two-fold higher than in the binary setting. This indicates that the comparative structure effectively leverages the probabilistic ranking capabilities of models to reduce classification noise more robustly than direct verification.

% \textbf{(3) MCQ provides stronger discriminative power than BCQ.} The random baseline represents the theoretical probability (50.00\% for BCQ and 25.00\% for MCQ), and the empirical validation results are provided in the Appendix \ref{suppl_sec:emp_random}. While the absolute accuracy scores are naturally lower in the multiple-choice setting, the relative improvement over the baseline is substantially higher. Specifically, the performance improvement across all models is approximately two-fold in the MCQ setting, which is significantly more pronounced than the gains observed in BCQ. This indicates that the multiple-choice format effectively filters out false positives derived from random guessing and provides a more robust metric for evaluating the true discriminative knowledge of the models.

\subsection{Impact of Thinking Process}

\begin{table}[]
\centering
\resizebox{0.8\columnwidth}{!}{%
\begin{tabular}{@{}cccc@{}}
\toprule
\textbf{model} & \textbf{Thinking} & \textbf{BCQ} & \textbf{MCQ} \\ \midrule
\multirow{2}{*}{GPT-5.1} & \yes & 60.77 & 52.30 \\
 & \no & 58.03 & 43.97 \\ \midrule
\multirow{2}{*}{Qwen3-Max} & \yes & 56.53 & 45.33 \\
 & \no & 54.57 & 40.80 \\ \bottomrule
\end{tabular}%
}
\caption{Performance comparison of LLMs with and without the thinking process on GenomeQA. The table reports the accuracy percentages on BCQ and MCQ.}
\label{tab:exp_thinking}
\end{table}

To examine the benefits of the explicit reasoning process, we evaluate selected LLMs equipped with the thinking mode. The goal is to assess their ability to perform step-by-step deductions and improve performance on complex genome tasks. We compare the performance of GPT-5.1 and Qwen3-Max by enabling and disabling their thinking features across both binary and multiple-choice settings. As shown in Table \ref{tab:exp_thinking}, the integration of the thinking process leads to consistent performance gains. Specifically, GPT-5.1 achieves a significant improvement in the multiple-choice setting, where the accuracy increases from 43.97\% to 52.30\%. This represents a notable enhancement in its ability to filter out distractors. Qwen3-Max also exhibits improvements, although the margins are smaller compared to GPT-5.1. For instance, its binary classification accuracy rises from 54.57\% to 56.53\%. The disparity in gains between the two models suggests that the effectiveness of the thinking mode depends heavily on the underlying domain knowledge of the base model. These results underscore the importance of enabling thinking capabilities to handle the intricate reasoning required for genome sequence analysis.

% \subsection{Impact of Multi-Step Reasoning}

% To investigate the challenge of indirect inference, we evaluate the impact of multi-step reasoning on model performance. We focus on the TFBS Prediction task, specifically selecting questions related to the 3D genome structure. The goal is to determine whether the performance bottleneck stems from pattern recognition capabilities or the difficulty of multi-step target inference. In the direct recognition setting, we explicitly provide the specific TF or feature target in the prompt, effectively removing the multi-step reasoning requirement. As shown in Table \ref{tab:inten_injection}, reducing the reasoning complexity results in performance improvements across all models. For instance, in the multiple-choice setting, Claude-Sonnet-4.5 and GPT-5.1 improves its accuracy from 27.11\% to 63.86\%. Similarly, Gemini3-Pro achieves a substantial gain, rising from 44.58\% to 67.47\% in the same setting. Performance under the original multi-step reasoning condition often hovers near the random baseline, whereas the direct recognition setting restores competitive accuracy. This indicates that models possess the underlying knowledge to identify genome features but struggle significantly to bridge the reasoning gap when the target must be inferred from indirect clues.

\subsection{Impact of Implicit Target Inference}

We design a controlled comparison to evaluate model performance on questions that require an additional inference step beyond direct sequence recognition. We focus on CTCF-related instances in the TFBS Prediction task. CTCF is a canonical transcription factor whose binding sites are strongly associated with higher-order chromatin organization, including chromatin loops and topologically associating domains (TADs). As a result, questions about 3D genome structure can implicitly point to CTCF, which must then be linked back to sequence-level evidence. Specifically, we construct two question variants over the same underlying sequence set. In the \emph{direct} setting, the question explicitly names the target (CTCF) and asks whether the sequence contains a CTCF binding site, which can be answered by direct pattern recognition. In the \emph{indirect} setting, the question does not mention CTCF and instead asks whether the sequence is associated with the formation of chromatin loops or TAD boundaries. Answering this variant requires a multi-step mapping: (i) infer the relevant regulatory factor implied by the functional description (CTCF), and (ii) evaluate whether the input sequence contains sequence patterns consistent with that factor. As shown in Table~\ref{tab:inten_injection}, making the target explicit substantially improves accuracy across models. For example, in the multiple-choice setting, Claude-Sonnet-4.5 and GPT-5.1 increase from 27.11\% to 63.86\%, and Gemini-3-Pro increases from 44.58\% to 67.47\%. In contrast, performance in the inference setting is often close to the random baseline. These results suggest that the additional target-inference step is a major source of difficulty in this setting.

\begin{table}[]
\centering
\resizebox{\columnwidth}{!}{%
\begin{tabular}{@{}ccccc@{}}
\toprule
\multirow{2}{*}{\textbf{Model}} & \multicolumn{2}{c}{\textbf{BCQ}} & \multicolumn{2}{c}{\textbf{MCQ}} \\ \cmidrule(l){2-5} 
 & \textbf{w/} & \textbf{w/o} & \textbf{w/} & \textbf{w/o} \\ \midrule
Claude-Sonnet-4.5 & 68.07 & 54.22 & 63.86 & 27.11 \\
GPT-5.1 & 68.07 & 52.41 & 63.86 & 27.11 \\
Gemini-3-Pro & 74.70 & 60.24 & 67.47 & 44.58 \\
Grok-4.1 & 61.45 & 54.82 & 58.43 & 27.11 \\
Llama-4 & 58.43 & 50.00 & 37.95 & 24.10 \\
Qwen3-Max & 67.47 & 45.78 & 67.45 & 24.10 \\ \bottomrule
\end{tabular}%
}
\caption{Impact of implicit target inference on TFBS prediction task. The table reports accuracy percentages where w/ denotes  the target (CTCF) is explicitly named, allowing direct pattern recognition and w/o denotes the implied target setting.}
% \caption{Impact of multi-step reasoning on model performance for the TFBS Prediction task. The table reports accuracy percentages where w/ denotes the direct recognition setting with explicit targets and w/o denotes the original multi-step reasoning setting.}
\label{tab:inten_injection}
\end{table}

\subsection{Failure Case Study}

\begin{figure}[h]
  \centering
  \includegraphics[width=0.6\linewidth]{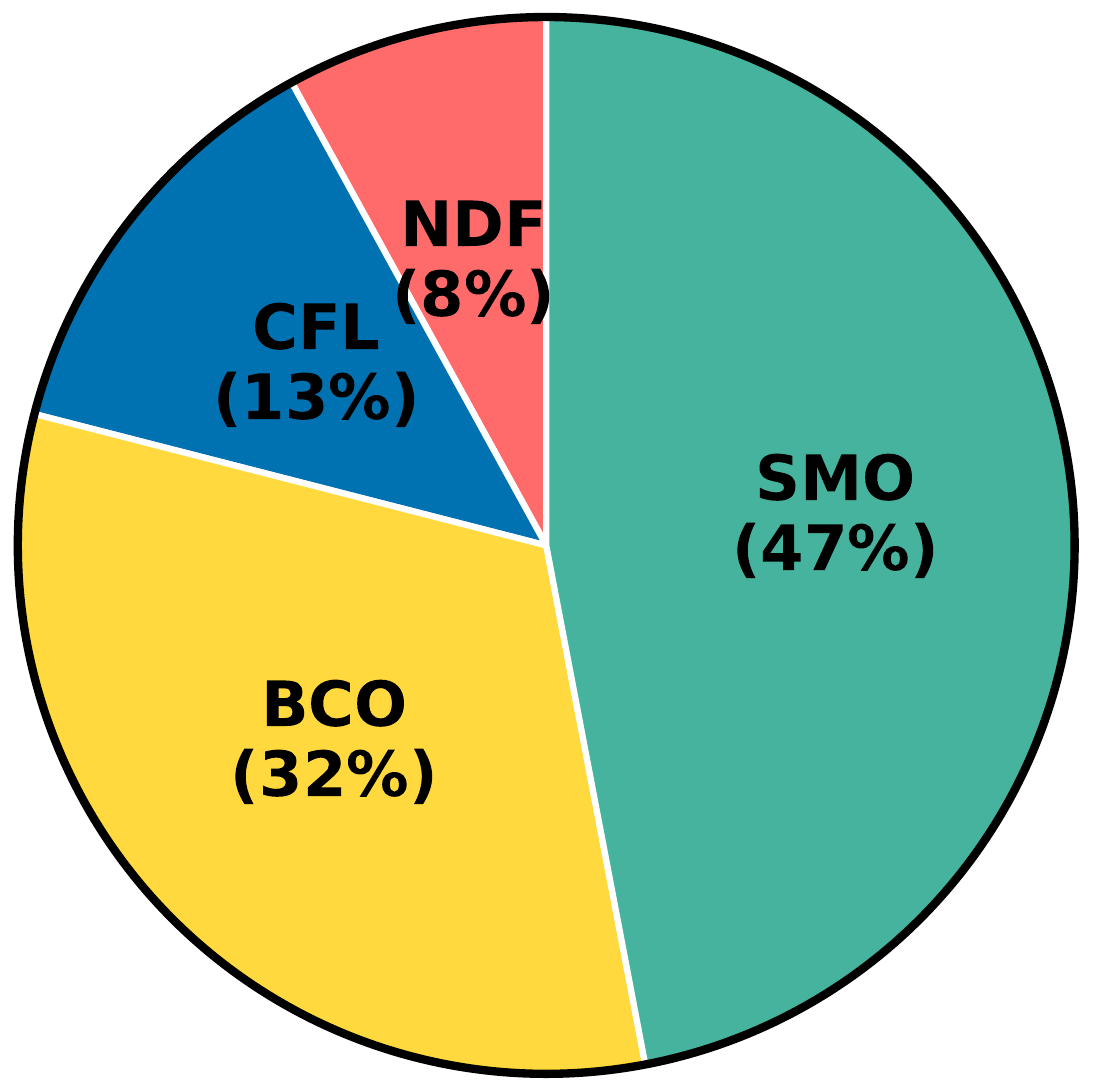}
  \caption{Distribution of failed cases.}
  \label{fig:fail_cls}
\end{figure}

To gain a deeper understanding of the limitations of current LLMs in genome analysis, we analyze 200 error samples produced by Gemini-3-Pro. As shown in Figure~\ref{fig:fail_cls}, our qualitative analysis categorizes these failures into four distinct types, revealing the cognitive disconnect between general LLM and the rigorous requirements of genomics. Details are provided in Appendix \ref{suppl_sec:case_study}.

\textbf{Sequence Motif Over-reliance (SMO, 47\%).} Failures occur when models rely on general sequence elements while neglecting specific details. For instance, in Histone Mark Prediction, the model incorrectly classifies an open \textit{Alu} repeat as closed \citep{alu_element}. It simply applies the general rule that transposable elements are repressed and overlooks the high GC content of this specific element.

\textbf{Base Composition Over-reliance (BCO, 32\%).} Failures occur when models rely on statistical summaries while ignoring structural patterns. For instance, in Taxonomic Classification, the model incorrectly identifies a virus as a prokaryote. It uses the high GC content as a shortcut and ignores the specific gene organization that actually points to a virus.

\textbf{Character Fidelity Loss (CFL, 13\%).} Models frequently lose character-level fidelity in long sequences, leading to the fabrication of non-existent sub-sequences to support their claims. In Enhancer and Promoter Identification, the model hallucinates a specific motif sequence as evidence that does not actually exist within the input sequence.

\textbf{Noise Distinction Failure (NDF, 8\%).} Failures occur when models fail to recognize meaningless patterns in shuffled negative samples. For instance, in Splice Site Identification, the model analyzes a randomized control sequence. It fails to detect the random nature of the input and performs a pseudo-analysis to incorrectly classify it as a splice site.

\section{Conclusion}

We introduce GenomeQA, a benchmark designed to support systematic evaluation of general-purpose large language models on sequence-based genome inference tasks. By curating biologically grounded tasks across multiple genomic contexts and adopting standardized question formats, GenomeQA provides a controlled evaluation setting for analyzing model performance on raw DNA sequences. Our experimental results show that current LLMs can leverage certain local sequence signals but exhibit substantial performance variation across tasks, particularly when targets are implicit or require additional inference steps. These findings highlight both the current limitations of general-purpose LLMs in this setting and the need for more reliable evaluation tools. We hope GenomeQA will serve as a useful diagnostic resource for future research on genome-aware language modeling and the integration of LLMs with genomic data.

\section*{Limitations}

We acknowledge two primary limitations of this work. \textbf{(1) Benchmark scale.} Evaluating multiple frontier LLMs under thinking-enabled settings incurs substantial computational cost. For example, Claude-Sonnet requires on the order of minutes to process a single sample, and Qwen3-Max can take even longer. To balance inference cost with task coverage, we curate a high-quality but moderately sized dataset. As a result, while GenomeQA is suitable for systematic evaluation, its current scale is not intended for full-parameter fine-tuning of large models. \textbf{(2) Task scope.} GenomeQA focuses on foundational sequence-based tasks such as motif recognition and coarse taxonomic classification. More complex biological problems, including variant effect prediction, and gene expression modeling, are not covered. These tasks typically require much longer sequences that exceed current context window limits and depend on additional modalities such as chromatin accessibility, histone modifications, and three-dimensional genome structure. Extending GenomeQA to incorporate such multi-omics signals is an important direction for future work.

% Second, the scope of genome tasks. While GenomeQA encompasses six diverse tasks covering regulatory elements, taxonomy, and modifications, it does not represent the full breadth of genome analysis. Complex structural tasks, such as 3D genome folding prediction, variant calling from raw reads, or long-range structural variation detection, are currently beyond the scope of this benchmark due to the context window limitations and the difficulty of representing such multi-dimensional data in a text-only format. Future iterations of GenomeQA aims to expand into these high-dimensional areas as model context windows continue to grow.

% Third, the opacity of training data in proprietary models. A significant portion of our evaluation relies on closed-source models (e.g., Gemini-3-Pro, GPT-5.1). Due to the opaque nature of their training corpora, we cannot rigorously verify the extent of data contamination—specifically, whether parts of our test sequences were present in their pre-training data. Although we selected specific genome regions and designed indirect questioning formats to mitigate memorization, the possibility of data leakage remains a challenge inherent to benchmarking closed-source foundation models.

\section*{Ethical considerations}

Our GenomeQA benchmark is built upon publicly available genome databases and established biological datasets, which are open for academic and research purposes. Additionally, we have rigorously reviewed the benchmark to ensure strict adherence to ethical guidelines. The dataset consists of de-identified public reference sequences and contains no personally identifiable information or private human genetic data. We have also verified that the benchmark focuses solely on fundamental biological understanding and contains no content related to biosecurity risks, harmful pathogens, or inappropriate material.

\bibliography{custom}

\newpage
\appendix

\section{Details for GenomeQA construction}
\subsection{Question Templates}
\label{sec:question_template}

As described in Section \ref{sec:design_principles}, we formulate two distinct question types: Binary Choice Questions (BCQ) and Multiple Choice Questions (MCQ). For each type, we curate a set of templates that were generally applied consistently across most tasks. While the template structure remains largely consistent across most tasks, we introduce specialized templates for the Splice Site Prediction, Histone Mark Prediction, and Transcription Factor Binding Site (TFBS) Prediction tasks to slightly increase the complexity of the tasks. Comprehensive examples of these templates are presented in Table \ref{tab:question_temp}.

% Please add the following required packages to your document preamble:
% \usepackage{booktabs}
% \usepackage{multirow}
% \usepackage{graphicx}
\begin{table*}[]
\centering
\resizebox{\textwidth}{!}{%
\begin{tabular}{@{}cll@{}}
\toprule
\textbf{Type} & \multicolumn{1}{c}{\textbf{Examples}} & \multicolumn{1}{c}{\textbf{Note}} \\ \midrule
\multirow{4}{*}{\textbf{BCQ}} & Does the Human DNA sequence \{seq\} contain any \{target\}? & \{seq\} represents a sequence, \{target\} is the query label \\
 & Does this Human DNA sequence contain authentic \{target\}? Sequence: \{seq\} & Specially used in Splice-Site Prediction \\
 & Consider the chromatin state in K562 cells. Is this Human DNA sequence located in an \{state\} region? Sequence: \{seq\} & Specially used in Histone Mark Prediction \\
 & We are looking for a sequence bound by the master regulator of chromatin looping and insulation. Is this sequence a match? \{seq\} & Specially used in TFBS prediction \\ \midrule
\multirow{4}{*}{\textbf{MCQ}} & Which of the following Human DNA sequences contains a \{target\}? & \{target\} is the query label \\
 & There is an exception in this group. While three sequences contain \{target1\}, one actually harbors a \{target2\}. Can you find it? & \{target1\} and \{target2\} represent query labels \\
 & Determine the chromatin accessibility state of this K562 sequence: \{seq\} & Specially used in Histone Mark Prediction \\
 & Which of the following Human DNA sequence is associated with the formation of chromatin loops and TAD boundaries? & Specially used in TFBS prediction \\ \bottomrule
\end{tabular}%
}
\caption{Question Templates in GenomeQA.}
\label{tab:question_temp}
\end{table*}

\begin{figure*}[htbp]
  \centering
  \includegraphics[width=\textwidth]{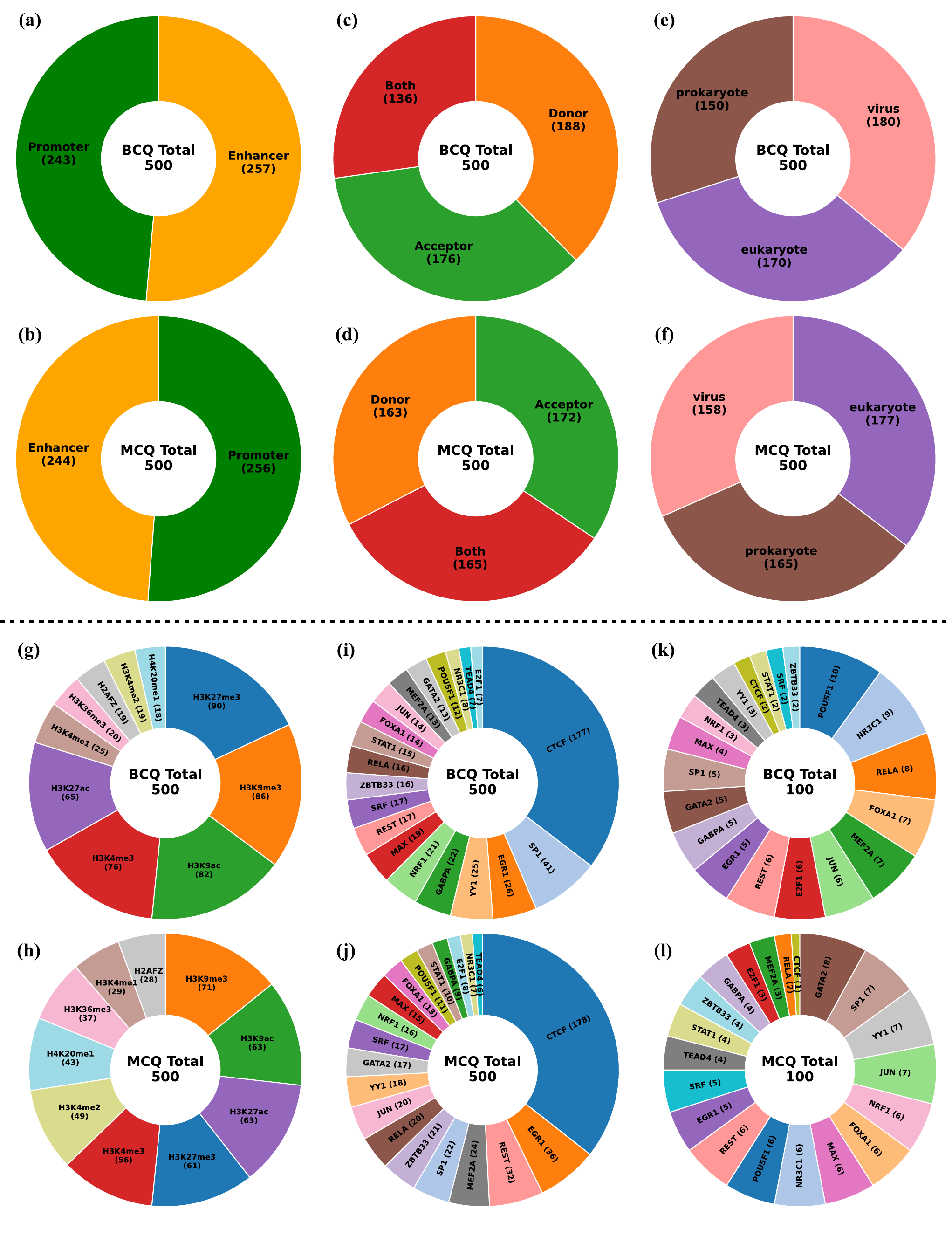}
  \caption{Label statistics for all tasks. The charts display the distribution of question types for BCQ and MCQ. (a)-(b): Enhancer and Promoter Identification. (c)-(d): Splice Site Identification. (e)-(f): Taxonomic Classification. (g)-(h): Histone Mark Prediction. (i)-(j): TFBS Prediction. (k)-(l): TF Motif Prediction.}
  \label{fig:suppl_all_task}
\end{figure*}

\subsection{Enhancer and Promoter Identification}
\label{sec:suppl_task1}
\subsubsection{Label Sets and Entity List}

For the Enhancer and Promoter Identification task, the label set is comprised of enhancer and promoter.

\subsubsection{Label Statistics}

% \begin{figure}[htbp]
%   \centering
%   \includegraphics[width=0.8\linewidth]{latex/suppl/suppl_task1.pdf}
%   \caption{Label statistics for the Enhancer and Promoter Identification task. The charts display the distribution of label occurrences in \textbf{(a)} BCQ and \textbf{(b)} MCQ.}
%   \label{fig:suppl_task1}
% \end{figure}

In this task, each sequence corresponds to a single, mutually exclusive label. We analyze the distribution of question types to verify dataset balance, as described in Figure \ref{fig:suppl_all_task}(a) and Figure \ref{fig:suppl_all_task}(b). In the Binary Choice Questions, the distribution is nearly symmetrical with 257 questions targeting enhancers and 243 targeting promoters. Similarly, the Multiple Choice Questions maintain this balanced approach featuring 244 enhancer-related queries and 256 promoter-related queries. This uniform distribution ensures that the model is evaluated equally on its ability to identify both regulatory elements.

\subsection{Splice Site Identification}
\label{sec:suppl_task2}
\subsubsection{Label Sets and Entity List}

For the Splice Site Identification task, the label set is comprised of acceptor, donor and both patterns. 

\subsubsection{Label Statistics}

In this task, each sequence corresponds to a single splice site category. In the Binary Choice Questions, the distribution remains relatively comparable across the three classes, with 176 questions targeting acceptors, 188 targeting donors, and 136 focusing on dual-site sequences. This balance extends to the Multiple Choice Questions, which feature an almost perfect split: 172 acceptor, 163 donor, and 165 both-related queries. Such a distribution guarantees that the performance metrics reflect a holistic understanding of all splice site configurations rather than a bias toward a specific type. The details are provided in Figure \ref{fig:suppl_all_task}(c) and Figure \ref{fig:suppl_all_task}(d).

% \begin{figure}[htbp]
%   \centering
%   \includegraphics[width=0.8\linewidth]{latex/suppl/suppl_task2.pdf}
%   \caption{Label statistics for the Splice Site Identification task. The charts display the distribution of question types for \textbf{(a)} BCQ and \textbf{(b)} MCQ.}
%   \label{fig:suppl_task2}
% \end{figure}

\subsection{Taxonomic Classification}
\label{sec:suppl_task3}

\subsubsection{Label Sets and Entity List}

In this task, the label set encompasses Eukaryote, Prokaryote, and Virus categories. We source eukaryotic sequences from Homo sapiens, Mus musculus, and Sus scrofa. For Prokaryotes, we select three common bacterial species. To address the shorter length of viral DNA, we utilize 9 viral species to ensure a sufficient number of candidate samples. Table \ref{tab:suppl_task3_list} provides detailed specifications.

% Please add the following required packages to your document preamble:
% \usepackage{booktabs}
% \usepackage{multirow}
% \usepackage{graphicx}
\begin{table}[]
\centering
\resizebox{\columnwidth}{!}{%
\begin{tabular}{@{}ccc@{}}
\toprule
\textbf{Taxonomy} & \textbf{Species} & \textbf{NCBI Accession} \\ \midrule
\multirow{3}{*}{\textbf{Eukaryote}} & Home sapiens & GCF\_000001405.40 \\
 & Mus musculus & GCF\_000001635.27 \\
 & Sus scrofa & GCF\_000003025.6 \\ \midrule
\multirow{3}{*}{\textbf{Prokaryote}} & Escherichia coli & GCF\_000005845.2 \\
 & Staphylococcus aureus & GCF\_000013425.1 \\
 & Bacillus subtilis & GCF\_000009045.1 \\ \midrule
\multirow{10}{*}{\textbf{Virus}} & Bacteriophage phiX174 & NC\_001422.1 \\
 & Simian virus 40 & NC\_001669.1 \\
 & Parvovirus B19 & NC\_000883.2 \\
 & Human papillomavirus type 16 & NC\_001526.4 \\
 & Human adenovirus 5 & NC\_001405.1 \\
 & Enterobacteria phage lambda & NC\_001416.1 \\
 & Human alphaherpesvirus 1 & NC\_001806.2 \\
 & Pandoravirus salinus & NC\_022098.1 \\
 & Megavirus chilensis & NC\_016072.1 \\
 & Acanthamoeba polyphaga mimivirus & NC\_014649.1 \\ \bottomrule
\end{tabular}%
}
\caption{Species in Taxonomic Classification task.}
\label{tab:suppl_task3_list}
\end{table}

\subsubsection{Label Statistics}

In this task, each sequence corresponds to a single taxonomic category, classifying inputs as Eukaryote, Prokaryote, or Virus. We analyze the distribution of question types to verify dataset balance, as described in Figure \ref{fig:suppl_all_task}(e) and Figure \ref{fig:suppl_all_task}(f). In the Binary Choice Questions, the dataset exhibits a well-proportioned composition across the three domains, with 170 questions targeting eukaryotes, 150 targeting prokaryotes, and 180 focusing on viral sequences. This balanced structure is mirrored in the Multiple Choice Questions, which feature a comparable distribution: 177 eukaryote, 165 prokaryote, and 158 virus-related queries. Such uniform coverage ensures that the model is rigorously evaluated on its ability to distinguish genome signatures across diverse biological domains without bias.

% \begin{figure}[htbp]
%   \centering
%   \includegraphics[width=0.8\linewidth]{latex/suppl/suppl_task3.pdf}
%   \caption{Label statistics for the Taxonomic Classification task. The charts display the distribution of question types for \textbf{(a)} BCQ and \textbf{(b)} MCQ.}
%   \label{fig:suppl_task3}
% \end{figure}

\subsection{Histone Mark Prediction}
\label{sec:suppl_task4}
\subsubsection{Label Sets and Entity List}

We utilize all 10 histone marks from the downstream Nucleotide Transformer dataset for the recognition task. Building on this, we select five histone marks to construct the chromatin accessibility task, consisting of three marks from open regions and two from closed regions. The "Undefined" represents the normal usage. Table \ref{tab:suppl_task4_list} contains detailed information.

% Please add the following required packages to your document preamble:
% \usepackage{booktabs}
% \usepackage{multirow}
% \usepackage{graphicx}
\begin{table}[]
\centering
\resizebox{\columnwidth}{!}{%
\begin{tabular}{@{}cc@{}}
\toprule
\textbf{Chrom. State} & \textbf{Selected Histone Marks} \\ \midrule
\textbf{Open} & H3K4me3, H3K9ac, H3K27ac \\
\textbf{Close} & H3K9me3, H3K27me3 \\
\multirow{2}{*}{\textbf{Undefined}} & H3K27me3, H2AFZ, H3K4me1 \\
 & H3K4me2, H3K36me3, H4K20me1 \\ \bottomrule
\end{tabular}%
}
\caption{Histone marks in the Histone Mark Prediction task. Chrom. State indicates the chromatin state.}
\label{tab:suppl_task4_list}
\end{table}

\subsubsection{Label Statistics}

In this task, each sequence corresponds to a single histone modification mark. The detailed information are provided in Figure \ref{fig:suppl_all_task}(g) and Figure \ref{fig:suppl_all_task}(h). Unlike the strictly uniform distributions in other tasks, the data here reveals an intentional skew in both Binary and Multiple Choice Questions. This deviation arises from the introduction of a Chromatin Accessibility reasoning dimension. To construct questions specifically focusing on open versus closed chromatin states, we prioritized a subset of five markers: three associated with open regions (H3K4me3, H3K27ac, H3K9ac) and two with closed regions (H3K9me3, H3K27me3). Consequently, questions targeting these five labels appear significantly more frequently than those targeting the remaining markers in the set, reflecting a design choice to stress-test the model's understanding of chromatin structure.

% \begin{figure}[htbp]
%   \centering
%   \includegraphics[width=\linewidth]{latex/suppl/suppl_task4.pdf}
%   \caption{Label statistics for the Histone Mark Prediction task. The charts display the distribution of question types (target labels) across the 10 histone marks for \textbf{(a)} BCQ and \textbf{(b)} MCQ. The visible unevenness reflects the deliberate oversampling of five specific markers to facilitate questions regarding chromatin accessibility.}
%   \label{fig:suppl_task4}
% \end{figure}

\subsection{TFBS Prediction}
\label{sec:suppl_task5}
\subsubsection{Label Sets and Entity List}

\begin{figure}[htbp]
  \centering
  \includegraphics[width=\linewidth]{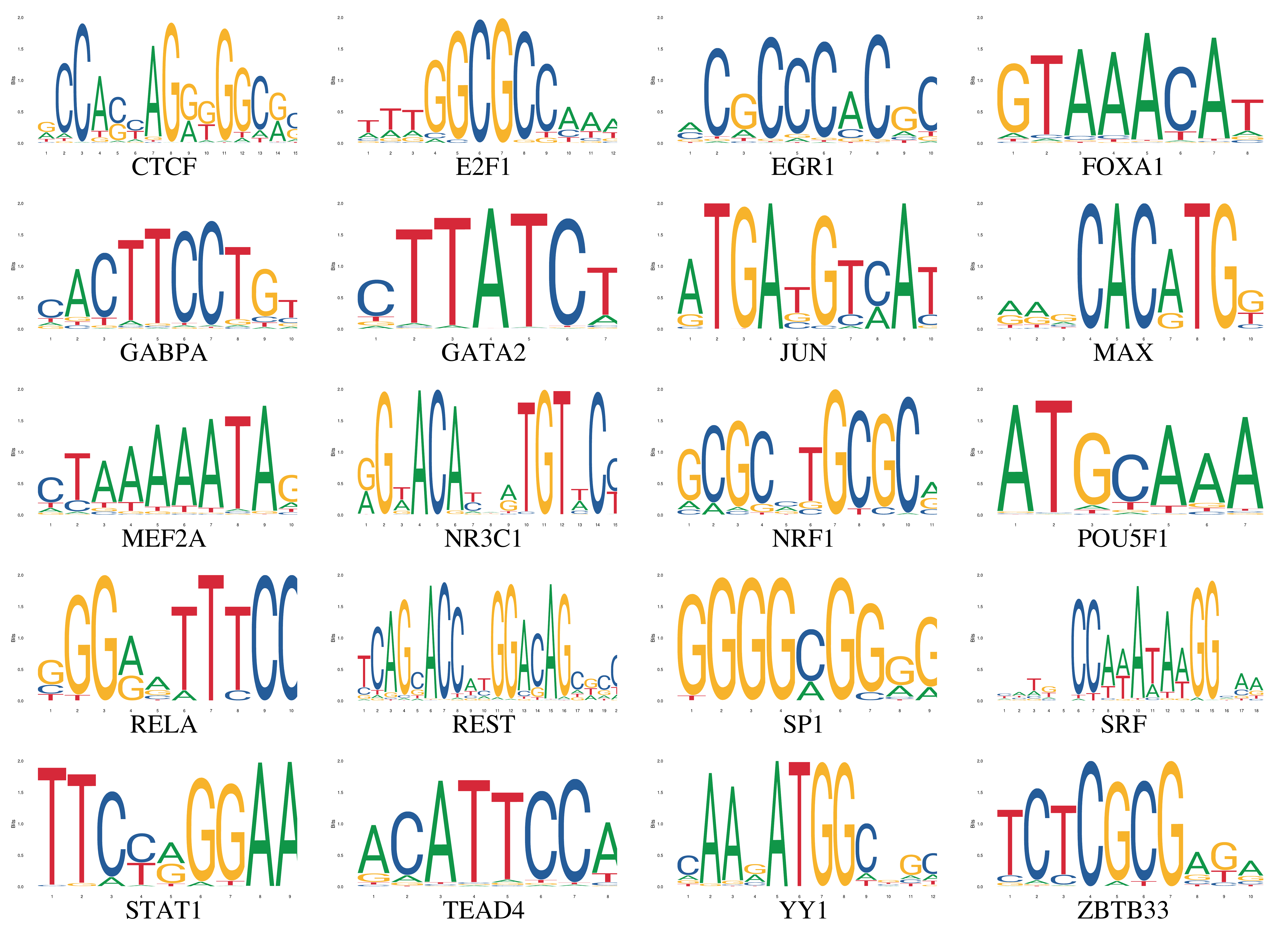}
  \caption{Motifs in TFBS Prediction.}
  \label{fig:suppl_task5_2}
\end{figure}

As described in Section \ref{sec:task_families}, we select 20 transcription factors with distinct motif patterns from the JASPAR database, as Figure \ref{fig:suppl_task5_2} illustrates. On this basis, we choose six factors to construct the 3D genome structure correlation task. Among these, CTCF associates with 3D architecture through its role in forming TADs and loops. The label Undefined indicates standard usage without contributing to additional problem dimensions. Table \ref{tab:suppl_task5_list} provides detailed specifications.

% Please add the following required packages to your document preamble:
% \usepackage{booktabs}
% \usepackage{multirow}
% \usepackage{graphicx}
\begin{table}[]
\centering
\resizebox{\columnwidth}{!}{%
\begin{tabular}{@{}cc@{}}
\toprule
\textbf{3D Genome} & \textbf{Selected TFs} \\ \midrule
\textbf{Related} & CTCF \\
\textbf{Unrelated} & SP1, GABPA, E2F1, TEAD4, NRF1 \\
\multirow{2}{*}{\textbf{Undefined}} & REST, EGR1, YY1, ZBTB33, FOXA1, JUN, MAX \\
 & SRF, MEF2A, RELA, STAT1, NR3C1, POU5F1, GATA2 \\ \bottomrule
\end{tabular}%
}
\caption{TFs in TFBS Prediction.}
\label{tab:suppl_task5_list}
\end{table}

\subsubsection{Label Statistics}

In this task, while genome sequences naturally contain multiple binding motifs (multi-label), each question interrogates the presence of a specific target Transcription Factor selected from a set of 20. The details are provided in Figure \ref{fig:suppl_all_task}(i) and Figure \ref{fig:suppl_all_task}(j). Similar to the Histone Mark task, the data exhibits a deliberate distributional skew across both Binary and Multiple Choice Questions. To evaluate the model's understanding of chromatin architecture, we significantly oversampled six specific TFs: CTCF, a key architectural protein facilitating TADs and loops, and SP1, GABPA, E2F1, TEAD4, and NRF1. Consequently, questions targeting these six factors dominate the statistics compared to the remaining 14 labels, prioritizing the assessment of structural regulatory logic over a uniform distribution.

% \begin{figure}[htbp]
%   \centering
%   \includegraphics[width=\linewidth]{latex/suppl/suppl_task5.pdf}
%   \caption{Label statistics for the TFBS Prediction task. The charts display the distribution of question types (target labels) across the 20 Transcription Factors for \textbf{(a)} BCQ and \textbf{(b)} MCQ. The distinct unevenness reflects the intentional oversampling of six specific TFs to facilitate questions regarding 3D genome structure and architectural proteins.}
%   \label{fig:suppl_task5}
% \end{figure}

\subsection{TF Motif Identification}
\label{sec:suppl_task6}
\subsubsection{Label Sets and Entity List}

We use the same label set as in the TFBS task. In this task, we do not differentiate between whether a factor is associated with 3D genome structure.

\subsubsection{Label Statistics}

In this task, each sequence corresponds to a single dominant motif selected from a set of 20 transcription factors. The statistics are shown in Figure \ref{fig:suppl_all_task}(k) and Figure \ref{fig:suppl_all_task}(l). In the Binary Choice Questions, the sampling is uniformly distributed across the 20 categories, ensuring no single motif dominates the evaluation. Similarly, the Multiple Choice Questions maintain this balanced approach, featuring an equitable distribution of target queries across the entire label set. This uniform coverage guarantees that the model’s ability to recognize short consensus patterns is tested fairly across diverse biological contexts without class bias.

% \begin{figure}[htbp]
%   \centering
%   \includegraphics[width=\linewidth]{latex/suppl/suppl_task6.pdf}
%   \caption{Label statistics for the TF Motif Prediction task. The charts display the distribution of question types (target labels) across the 20 motif categories for \textbf{(a)} BCQ and \textbf{(b)} MCQ.}
%   \label{fig:suppl_task6}
% \end{figure}

\section{System Prompt Optimization}
\label{sec:prompt_ablation}

To ensure that the evaluation accurately reflects the genome reasoning capabilities of Large Language Models (LLMs) rather than their sensitivity to prompt engineering, we conduct an ablation study to derive an optimized system prompt. This prompt aims to maximize the models' ability to interpret DNA sequences by providing clear domain constraints and reasoning protocols.

\subsection{Development Set Construction}

To balance computational cost with statistical representativeness, we construct a small-scale development set comprising 90 samples. We employ a stratified sampling strategy by randomly selecting three samples for each unique combination of problem type, task category, and specific problem dimension. This selection spans binary and multiple-choice formats across tasks such as TFBS prediction and splice site identification, including specific dimensions like 3D genome structure correlation. This compact yet diverse dataset serves as the testbed for iterative prompt refinement.

\subsection{Optimization Strategy}

The optimization process follows a two-stage iteration starting with a baseline prompt generated by Gemini 3 Pro, as Figure \ref{fig:base_prompt} illustrates. This initial text provides a general persona of a computational biology expert and standard instructions to analyze sequences. We conduct manual error analysis on the model outputs and observe that models frequently hallucinate constraints or refuse to answer due to a perceived lack of evidence. To address this, we manually refine the text into the Optimized System Prompt presented in Figure \ref{fig:optimized_prompt}. Key enhancements include explicitly instructing models to choose the most likely option based on probabilistic signals, defining six specific biological domains to narrow the search space, and mandating a structured analysis of motif organization. This final requirement specifically aids in distinguishing real sequences from dinucleotide-preserved shuffled controls.

\begin{figure*}[h] % 或者 figure [t] 如果你想把它做成单栏
    \centering
    % 使用灰色边框，代表 Baseline
    \begin{promptbox}[colframe=gray!50!black]{Base Prompt}
        You are an expert in Computational Biology and Bioinformatics. Your task is to solve problems related to genome sequence analysis.
        
        \vspace{0.5em}
        You will be provided with a specific question and a set of options. The DNA sequences may appear in the question description or within the options themselves.
        
        \vspace{0.5em}
        Please follow these instructions:
        \begin{enumerate}[nosep, leftmargin=1.5em]
            \item Identify the specific biological task (e.g., distinguishing enhancer vs. promoter).
            \item Briefly analyze the key sequence features or biological context relevant to the question. Keep this analysis concise (1-3 sentences) to justify your choice.
            \item Select the most accurate option letter (e.g., A, B, C, D) .
        \end{enumerate}
        
        \vspace{0.5em}
        Strictly follow the following output format:\\
        \#\#\# Analysis\\
        {[}Concise analysis in 1-3 sentences{]}
        
        \vspace{0.5em}
        \#\#\# Answer\\
        Answer: {[}Option Letter{]}
    \end{promptbox}
    \caption{The Base Prompt used as a baseline. It contains only basic task descriptions and formatting instructions without domain-specific reasoning guidance.}
    \label{fig:base_prompt}
\end{figure*}

\begin{figure*}[t]
    \centering
    % 使用 Teal 色 (蓝绿色) 边框，看起来比较高级且区别于 Base
    \begin{promptbox}[colframe=teal!60!black]{Optimized Prompt}
        You are an expert in Computational Biology, Regulatory Genomics, and Bioinformatics. Your goal is to infer the correct answer by analyzing nucleotide sequences with probabilistic reasoning and biological domain knowledge, not by exploiting superficial text patterns or option positions.
        
        \vspace{0.5em}
        Questions implicitly belong to one of six domains (you must infer which):
        \begin{enumerate}[nosep, leftmargin=1.5em]
            \item Regulatory identity: enhancer vs promoter in human.
            \item Splice-site analysis: acceptor vs donor vs both, and distinguishing real sites from dinucleotide-preserved shuffled controls.
            \item Taxonomy: classifying sequences as eukaryote, prokaryote, or virus.
            \item Epigenetics: histone mark identity and chromatin state (open/accessible vs closed/repressive) in human K562 cells.
            \item TFBS prediction: locating binding sites of specific transcription factors within longer human sequences.
            \item Motif identification: recognizing short (6-20 bp) consensus patterns.
        \end{enumerate}
        
        \vspace{0.5em}
        Please adhere to the following Reasoning Protocol:
        \begin{enumerate}[leftmargin=1.5em, parsep=0.5em]
            \item \textbf{Infer task \& question type}\\
            From the wording, infer the biological domain and whether the question is: binary verification, multi-class label selection, sequence selection, odd-one-out, or real-vs-shuffled discrimination.
            
            \item \textbf{Analyze sequence patterns}\\
            Assume all sequences are in functional 5'->3' orientation unless stated otherwise. Do NOT perform reverse-complement analysis unless explicitly asked. Focus on informative patterns:
            \begin{itemize}[nosep, leftmargin=1em]
                \item Short motifs and motif-like patterns, their approximate positions and local context.
                \item Base composition, GC content, k-mer and codon-like usage, periodicity, and sequence complexity.
                \item Differences between options (for selection / odd-one-out), such as one sequence having a stronger or more coherent pattern than the others.
            \end{itemize}
            Treat any examples you know (e.g., canonical splice motifs, promoter features, TF consensus-like sites, species-specific composition) as *hints*, not as a closed list: you are encouraged to use any additional recurrent or statistically distinctive patterns you detect in the sequences, even if they are not classic textbook motifs.
            
            \item \textbf{Make a probabilistic decision}\\
            Do NOT require perfect consensus matches; biological signals are degenerate and noisy.
            \begin{itemize}[nosep, leftmargin=1em]
                \item Interpret strict verbs (``is this X?'', ``confirm'') as asking which label is *more likely*.
                \item For real vs dinucleotide-preserved shuffled/background comparisons, remember that low-level composition is similar; rely more on motif organization, local structure, and plausibly functional subpatterns.
                \item Do not hedge in your analysis; always choose the single most biologically plausible option.
            \end{itemize}
        \end{enumerate}
        
        \vspace{0.5em}
        Output rules:
        \begin{itemize}[nosep, leftmargin=1.5em]
            \item Base your decision strictly on nucleotide patterns and biological meaning of the labels, not on dataset biases or text-only artifacts.
            \item Do NOT copy or restate any full DNA sequence in your analysis; refer only to features (motifs, composition, relative differences, etc.).
            \item For all questions, there is exactly one best answer. Always pick a single option letter.
        \end{itemize}
        
        \vspace{0.5em}
        Output format (strictly follow):\\\\
        \#\#\# Analysis\\
        1-3 concise sentences explaining (i) the inferred task/domain and (ii) the key sequence-based or comparative reasoning that makes your chosen option most probable.\\
        % \vspace{0.3em}
        
        \#\#\# Answer\\
        Answer: {[}Single option letter, e.g., A / B / C / D{]}
    \end{promptbox}
    \caption{The Optimized Prompt designed for the LLM. It includes explicit role definition, domain constraints, a step-by-step reasoning protocol, and strict output formatting rules.}
    \label{fig:optimized_prompt}
\end{figure*}

\subsection{Performance Validation}

We evaluate the effectiveness of the optimized prompt against the baseline across six LLMs on the development set. As the ablation studies in Table \ref{tab:ablation_bcq} and Table \ref{tab:ablation_mcq} demonstrate, the optimized prompt consistently matches or outperforms the baseline version. Across all six evaluated models, the optimized prompt yields performance gains in at least three out of six tasks per model. The improvement is particularly notable in tasks requiring subtle pattern recognition, where the explicit instruction to avoid hedging forces models to leverage weak biological signals for decision-making. 
% Consequently, we adopt the optimized system prompt for the main experiments reported in this work.

% Please add the following required packages to your document preamble:
% \usepackage{booktabs}
% \usepackage{multirow}
% \usepackage{graphicx}
\begin{table*}[h]
\centering
\resizebox{\textwidth}{!}{%
\begin{tabular}{@{}cccccccccccccccc@{}}
\toprule
\multirow{3}{*}{\textbf{Model}} & \multicolumn{15}{c}{\textbf{BCQ (33 samples)}} \\ \cmidrule(l){2-16} 
 & \multicolumn{2}{c}{\textbf{\begin{tabular}[c]{@{}c@{}}Enhancer and Promoter\\ Identification\end{tabular}}} & \multicolumn{2}{c}{\textbf{\begin{tabular}[c]{@{}c@{}}Splice Site\\ Identification\end{tabular}}} & \multicolumn{2}{c}{\textbf{\begin{tabular}[c]{@{}c@{}}Taxonomic\\ Classification\end{tabular}}} & \multicolumn{2}{c}{\textbf{\begin{tabular}[c]{@{}c@{}}Histone Mark\\ Prediction\end{tabular}}} & \multicolumn{2}{c}{\textbf{\begin{tabular}[c]{@{}c@{}}TFBS\\ Prediction\end{tabular}}} & \multicolumn{2}{c}{\textbf{\begin{tabular}[c]{@{}c@{}}TF Motif\\ Prediction\end{tabular}}} & \multicolumn{2}{c}{\textbf{Avg.}} & \multirow{2}{*}{\textbf{\begin{tabular}[c]{@{}c@{}}No. Task\\ Wins\end{tabular}}} \\ \cmidrule(lr){2-15}
 & Base & Opt. & Base & Opt. & Base & Opt. & Base & Opt. & Base & Opt. & Base & Opt. & Base & Opt. &  \\ \midrule
\textbf{Claude-Sonnet-4.5} & 100.00 & 66.67 & 44.44 & 22.22 & 33.33 & \textbf{100.00} & 33.33 & \textbf{50.00} & 66.67 & 55.56 & 66.67 & \textbf{66.67} & 57.41 & \textbf{60.19} & 3 \\
\textbf{GPT-5.1} & 100.00 & \textbf{100.00} & 88.89 & 44.44 & 100.00 & 66.67 & 66.67 & \textbf{66.67} & 55.56 & \textbf{66.67} & 33.33 & \textbf{100.00} & 74.07 & \textbf{74.07} & 4 \\
\textbf{Gemini3-Pro} & 66.67 & \textbf{100.00} & 44.44 & \textbf{55.56} & 100.00 & \textbf{100.00} & 50.00 & 33.33 & 77.78 & \textbf{77.78} & 100.00 & \textbf{100.00} & 73.15 & \textbf{77.78} & 5 \\
\textbf{Grok-4.1} & 100.00 & \textbf{100.00} & 33.33 & \textbf{66.67} & 66.67 & \textbf{66.67} & 33.33 & \textbf{50.00} & 66.67 & \textbf{66.67} & 33.33 & \textbf{66.67} & 55.56 & \textbf{69.44} & 6 \\
\textbf{Llama4} & 66.67 & \textbf{66.67} & 55.56 & \textbf{55.56} & 66.67 & \textbf{66.67} & 83.33 & \textbf{83.33} & 77.78 & 66.67 & 0.00 & \textbf{66.67} & 58.33 & \textbf{67.59} & 5 \\
\textbf{Qwen3-Max} & 0.00 & \textbf{100.00} & 66.67 & \textbf{77.78} & 66.67 & \textbf{66.67} & 50.00 & \textbf{83.33} & 66.67 & 55.56 & 100.00 & 66.67 & 58.33 & \textbf{75.00} & 4 \\ \bottomrule
\end{tabular}%
}
\caption{Ablation study on Binary Choice Questions. Performance comparison between the initial and refined prompts on the 33-sample BCQ development set. Base denotes the Base System Prompt and Opt. denotes the Optimized System Prompt. Avg. represents the average accuracy across all six tasks. No. Task Wins indicates the number of tasks where the optimized prompt achieved equal or higher accuracy than the base prompt. Bold values highlight instances where the optimized prompt resulted in improved or equivalent performance compared to the baseline.}
\label{tab:ablation_bcq}
\end{table*}

% Please add the following required packages to your document preamble:
% \usepackage{booktabs}
% \usepackage{multirow}
% \usepackage{graphicx}
\begin{table*}[h]
\centering
\resizebox{\textwidth}{!}{%
\begin{tabular}{@{}cccccccccccccccc@{}}
\toprule
\multirow{3}{*}{\textbf{Model}} & \multicolumn{15}{c}{\textbf{MCQ (57 samples)}} \\ \cmidrule(l){2-16} 
 & \multicolumn{2}{c}{\textbf{\begin{tabular}[c]{@{}c@{}}Enhancer and Promoter\\ Identification\end{tabular}}} & \multicolumn{2}{c}{\textbf{\begin{tabular}[c]{@{}c@{}}Splice Site\\ Identification\end{tabular}}} & \multicolumn{2}{c}{\textbf{\begin{tabular}[c]{@{}c@{}}Taxonomic\\ Classification\end{tabular}}} & \multicolumn{2}{c}{\textbf{\begin{tabular}[c]{@{}c@{}}Histone Mark\\ Prediction\end{tabular}}} & \multicolumn{2}{c}{\textbf{\begin{tabular}[c]{@{}c@{}}TFBS\\ Prediction\end{tabular}}} & \multicolumn{2}{c}{\textbf{\begin{tabular}[c]{@{}c@{}}TF Motif\\ Prediction\end{tabular}}} & \multicolumn{2}{c}{\textbf{Avg.}} & \multirow{2}{*}{\textbf{\begin{tabular}[c]{@{}c@{}}No. Task\\ Wins\end{tabular}}} \\ \cmidrule(lr){2-15}
 & Base & Opt. & Base & Opt. & Base & Opt. & Base & Opt. & Base & Opt. & Base & Opt. & Base & Opt. &  \\ \midrule
\textbf{Claude-Sonnet-4.5} & 66.67 & \textbf{77.78} & 44.44 & 33.33 & 77.78 & \textbf{88.89} & 40.00 & \textbf{66.67} & 44.44 & \textbf{66.67} & 83.33 & \textbf{83.33} & 59.44 & \textbf{69.44} & 5 \\
\textbf{GPT-5.1} & 44.44 & \textbf{77.78} & 11.11 & \textbf{22.22} & 66.67 & \textbf{66.67} & 40.00 & \textbf{53.33} & 44.44 & \textbf{55.56} & 100.00 & 83.33 & 51.11 & \textbf{59.81} & 5 \\
\textbf{Gemini3-Pro} & 66.67 & 55.56 & 22.22 & \textbf{44.44} & 100.00 & 88.89 & 40.00 & \textbf{46.67} & 22.22 & \textbf{44.44} & 100.00 & \textbf{100.00} & 58.52 & \textbf{63.33} & 4 \\
\textbf{Grok-4.1} & 44.44 & \textbf{44.44} & 33.33 & 22.22 & 55.56 & 44.44 & 33.33 & \textbf{40.00} & 77.78 & 66.67 & 100.00 & \textbf{100.00} & 57.41 & 52.96 & 3 \\
\textbf{Llama4} & 33.33 & 11.11 & 11.11 & \textbf{11.11} & 77.78 & 55.56 & 46.67 & \textbf{46.67} & 55.56 & \textbf{66.67} & 66.67 & \textbf{83.33} & 48.52 & 45.74 & 4 \\
\textbf{Qwen3-Max} & 55.56 & 44.44 & 11.11 & \textbf{11.11} & 66.67 & \textbf{77.78} & 33.33 & \textbf{40.00} & 77.78 & 55.56 & 83.33 & \textbf{83.33} & 54.63 & 52.04 & 4 \\ \bottomrule
\end{tabular}%
}
\caption{Ablation study on Multiple Choice Questions. Performance comparison on the 57-sample MCQ development set. Base denotes the Base System Prompt and Opt. denotes the Optimized System Prompt. Avg. represents the average accuracy across all six tasks. No. Task Wins indicates the number of tasks where the optimized prompt achieved equal or higher accuracy than the base prompt. Bold formatting signifies that the optimized prompt achieved performance parity or improvement over the base prompt.}
\label{tab:ablation_mcq}
\end{table*}

\section{Empirical Baseline Verification}
\label{suppl_sec:emp_random}

To validate the reliability of using theoretical probability as a performance floor, we calculate empirical random baselines by simulating random guessing across all evaluation tasks. We compare these empirical values against the standard theoretical expectations of 50.00\% for Binary Choice Questions (BCQ) and 25.00\% for Multiple Choice Questions (MCQ).

As presented in Table \ref{tab:random_baseline}, the empirical results align closely with theoretical expectations. The average empirical accuracy across all tasks is 51.17\% for BCQ and 25.23\% for MCQ, exhibiting only negligible deviations from the theoretical values. The overall consistency confirms that the theoretical baselines serve as accurate and fair proxies for zero-knowledge performance in our main analysis.

% Please add the following required packages to your document preamble:
% \usepackage{graphicx}
\begin{table}[H]
\centering
\resizebox{\columnwidth}{!}{%
\begin{tabular}{ccc}
\toprule
\textbf{Task} & \textbf{BCQ} & \textbf{MCQ} \\ \midrule
\begin{tabular}[c]{@{}c@{}}Enhancer and Promoter\\ Identification\end{tabular} & 51.40 & 25.80 \\
Splice Site Identification & 52.00 & 24.00 \\
Taxonomic Classification & 50.80 & 28.20 \\
Histone Mark Prediction & 49.60 & 23.00 \\
TFBS Prediction & 49.20 & 22.40 \\
TF Motif Prediction & 54.00 & 28.00 \\ \midrule
Avg & 51.17 & 25.23 \\ \bottomrule
\end{tabular}%
}
\caption{Empirical random baseline accuracy (\%) across different tasks in GenomeQA.}
\label{tab:random_baseline}
\end{table}

\section{Case Study}
\label{suppl_sec:case_study}

\begin{figure*}[h]
  \centering
  \includegraphics[width=0.8\textwidth]{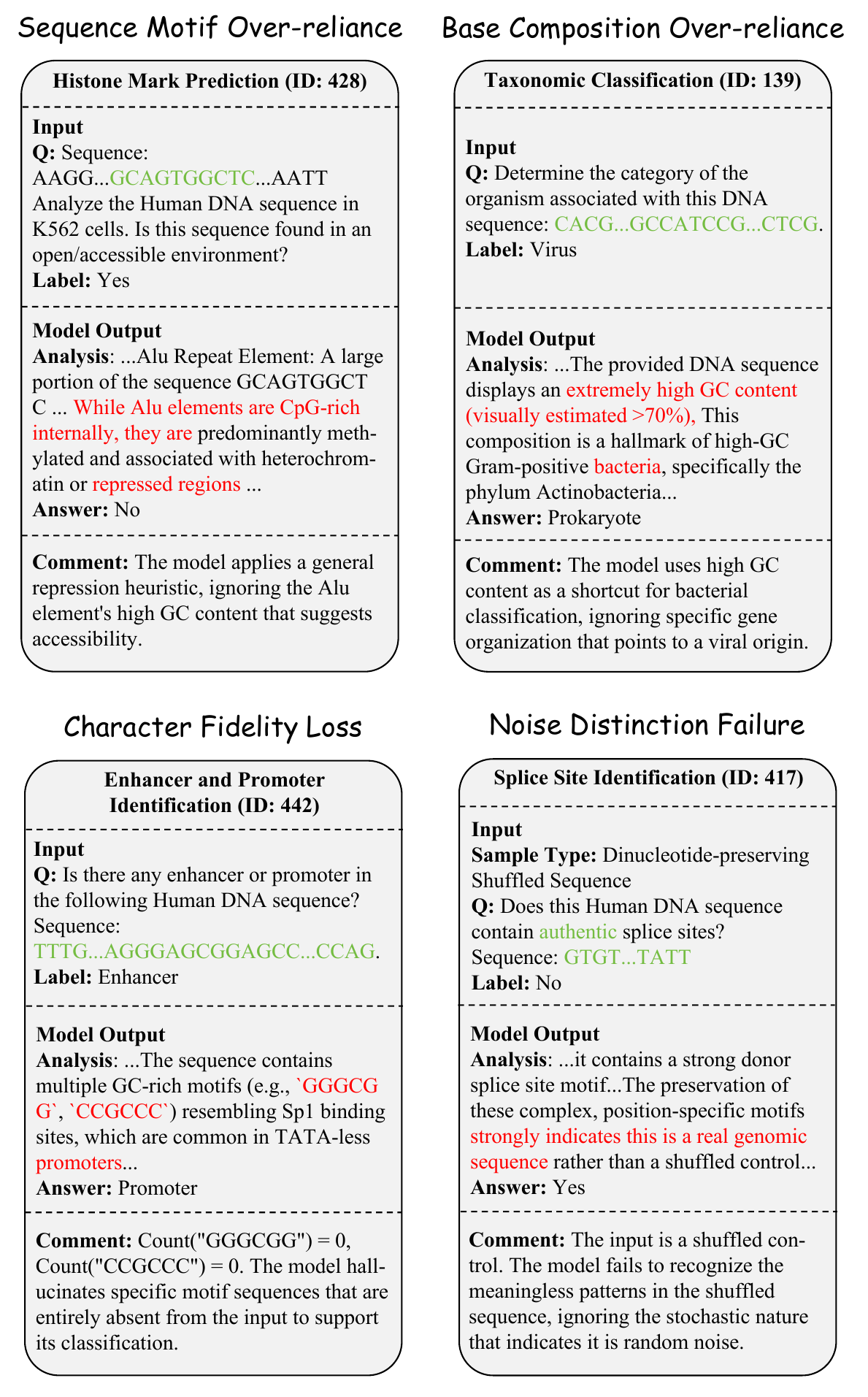}
  \caption{Case Study. Examples of failure modes in GenomeQA along with input details and model responses. Green: Ground truth labels, specific input characteristics provided by the benchmark. Red: Incorrect reasoning, or fabricated evidence generated by the model.}
  \label{fig:suppl_case_study}
\end{figure*}

Figure~\ref{fig:suppl_case_study} presents examples of four failure modes in genome analysis: Sequence Motif Over-reliance (SMO), Base Composition Over-reliance (BCO), Character Fidelity Loss (CFL), and Noise Distinction Failure (NDF). Each example includes the input context, model output, and an analysis of the error. These cases highlight distinct cognitive gaps in general LLMs: prioritizing general sequence elements over specific details (SMO), using statistical shortcuts like GC content while ignoring structural patterns (BCO), losing character-level fidelity to fabricate non-existent sub-sequences (CFL), and the sycophantic tendency to rationalize random noise as valid biological signals (NDF). These examples illustrate the systematic error patterns in current LLMs and underscore the need for domain-aligned reasoning capabilities.
% Figure \ref{fig:suppl_case_study} presents examples of four failure cases in genome analysis: Flawed Reasoning, Constraint Violation, Fact Hallucination, and Confirmation Bias. Each example includes the input context, model output, and an analysis of the error. The examples highlight distinct challenges encountered by general LLMs in this domain: reliance on superficial statistics rather than deep biological causalities (Flawed Reasoning), the tendency to override explicit negative constraints to manufacture positive answers (Constraint Violation), the loss of character-level fidelity leading to the fabrication of non-existent sub-sequences (Fact Hallucination), and sycophantic behavior that attempts to rationalize random noise as valid biological signals (Confirmation Bias). These examples illustrate the diverse error patterns in current LLMs and highlight the need for more robust, domain-aligned genome reasoning capabilities.

\end{document}